\documentclass[a4paper,12pt,twoside]{article}
\usepackage{amsfonts,graphicx}
\newcommand{\rnc}{\renewcommand}
\newcommand{\eqreset}{\setcounter{equation}{0}}
\rnc{\thesection}{\arabic{section}\setcounter{equation}{0}}
\rnc{\theequation}{\arabic{section}.\arabic{equation}}
\rnc{\thefootnote}{\alph{footnote}}

\begin{document}
\baselineskip=8mm
\title{
Thermodynamics of $osp(1|2)$ Integrable Spin Chain:
Finite Size Correction}
\author{Kazumitsu SAKAI$^{*}$  
\\ 
{\it Universit\"at Dortmund, 
Theoretische Physik I,} \\ 
{\it Otto-Hahn-Str. 4, D-44221 Dortmund, Germany} \\ and \\ 
Zengo TSUBOI$^{**}$ \\
{\it Institute of Physics,                                   
 University of Tokyo,} \\
{\it Komaba   3-8-1, Meguro-ku, Tokyo 153-8902, Japan}}
\date{}
\maketitle
\begin{abstract}
This note is a supplement to our previous papers: 
{\it Mod. Phys. Lett.} {\bf A14} (1999) 2427 
(math-ph/9911010); 
{\it Int. J. Mod. Phys.} {\bf A15} (2000) 2329 
(math-ph/9912014).

The thermodynamic Bethe ansatz (TBA) equation
for an integrable spin chain related to the Lie 
superalgebra $osp(1|2)$ is analyzed. 
The central charge determined by  low temperature
asymptotics of the specific heat can be expressed by the 
Rogers dilogarithmic function, and identified to be  
$1$. 
Solving the TBA equation numerically, 
we evaluate  the several thermodynamic 
quantities.
The excited state TBA equation is also discussed. 
\footnotetext{$^{*}$E-mail: sakai@printfix.physik.uni-dortmund.de}
\footnotetext{$^{**}$E-mail: tsuboi@gokutan.c.u-tokyo.ac.jp}
\footnotetext{$^{* \ **}$alphabetic order}
\end{abstract}
{\bf to appear in J. Phys. Soc. Jpn. Vol. 70-2}
\section{Introduction}
Recently, thermodynamics of quantum integrable spin chains 
related to Lie superalgebras received much attentions. 
In particular, several people 
\cite{Sch87,EK94,JKS98,Fr99,Sa00} 
studied thermodynamic Bethe ansatz (TBA) 
equations \cite{YY69} related to $sl(r|s)$. 
On the other hand, study on $osp(r|2s)$ case 
has begun only recently in refs. \cite{ST99,ST00} 
(cf. ref. \cite{T00}), 
in which we deal with the simplest $osp(1|2)$ model 
\cite{Kul86,BS88}.  
Namely, we have derived the $Y$-system from 
the $osp(1|2)$ version of the $T$-system 
\cite{T99}, and transformed it into 
the TBA equation and excited state TBA equation 
from the point of view of the quantum transfer matrix (QTM) 
method \cite{Kl92}. 
As for the largest eigenvalue sector of the 
dressed vacuum form (DVF), 
this TBA equation coincides  with the
one \cite{ST99} from the traditional
string hypothesis \cite{T71,G71}. 
On the other hand, the excited state TBA equation from 
the second largest eigenvalue sector, which 
characterizes the correlation length, 
is difficult to derive by the string hypothesis. 
The purpose of this paper is applications of
 our previous results \cite{ST99,ST00} 
to the calculation of physical quantities. 

%
The Hamiltonian \cite{Mar95} 
of the $osp(1|2)$ integrable spin chain, which we deal with 
in this paper is given by
\begin{eqnarray}
H=J\sum_{j=1}^{L}
 \left(
   P^{g}_{j, j+1}+\frac{2}{3}E_{j, j+1}
 \right) \label{hami}, 
\end{eqnarray}
where we assume the periodic boundary condition.
For precise definition of (\ref{hami}), see refs. \cite{Mar95,ST00}. 
In this paper, we consider the case $J=-1$, 
which corresponds to the antiferromagnetic regime.   

In section 2, we consider the finite temperature correction 
of the  DVF $T^{(1)}_{1}(v)$ from the $T$-system \cite{T99} 
and calculate the central charge from the low temperature 
asymptotics 
of the specific heat (see, refs. \cite{KP92,KNS94-2,BCN86,A86,BR89}). 
The central charge is expressed in terms of the Rogers dilogarithm
function,
and reproduces the conjecture \cite{Mar95} 
$c=1$ from the root density method. 
Moreover we solve the TBA equation numerically 
and evaluate the thermodynamic quantities
such as the free energy, the internal energy, 
the specific heat and the entropy, which are depicted in 
fig. \ref{fig}. 
To the author's knowledge, this is the first 
evaluations of the physical quantities of 
$osp(1|2)$ model at {\em finite} temperature. 

Above calculation corresponds to the largest eigenvalue 
sector of the DVF. 
In section 3 we comment on the excited state TBA equations, 
which have singularities from zeros of fused QTMs. 
For comparison, we also briefly mention a finite temperature correction 
from the traditional string hypothesis in Appendix. 
\section{Analysis of the TBA Equation}
In our previous papers \cite{ST99,ST00}, we 
proposed the TBA equation of the $osp(1|2)$ model
\begin{eqnarray}
\log Y_{m}(v)=
-\frac{\pi \beta \delta_{m 1}}{\cosh\pi v}+
K*\log
\left\{\frac{(1+Y_{m+1})(1+Y_{m-1})}
{1+Y_{m}^{-1}}
\right\}(v), 
\label{tba}
\end{eqnarray}
where $m\in{\mathbb Z}_{\ge 1}$, $Y_{0}(v):=0$, 
$\beta=1/T$ ($T$: temperature; in this paper, 
we set the Boltzmann constant to $1$.), 
$K(v)$ is the kernel defined by
\begin{equation}
K(v)=\frac{1}{2\cosh\pi v},
\label{kernel1}
\end{equation}
and  $A*B(v)$ denotes the convolution 
\begin{equation}
A*B(v)=\int_{-\infty}^{\infty}{\rm d}w
A(v-w)B(w).
\end{equation}
Through this TBA equation, the logarithm of the largest eigenvalue of the
QTM $T^{(1)}_1(v)$ (See, ref.\cite{ST00} for the definition of 
$T^{(1)}_{m}(v)$.) can be expressed as
\begin{equation}
\log T^{(1)}_1(0)=\beta\left( \frac{4\pi}{3\sqrt{3}}-1\right)+G*
\log(1+Y_1)(0),
\label{free1}
\end{equation}
and the free energy per site 
$f$ is written as
\begin{equation}
f=-T \log T^{(1)}_1(0). \label{free2}
\end{equation}
Here the kernel $G(v)$ is defined by
\begin{equation}
G(v)=
\frac{2 \sinh \frac{4\pi v}{3}}
{\sqrt{3}\sinh 2\pi v}.
\label{kernel2}
\end{equation}
Note that $-4\pi/(3\sqrt{3})+1=:E_{\rm gs}$ in
eq. (\ref{free1}) is the ground state energy
 \cite{Mar95} of (\ref{hami}).

Following ref. \cite{KP92}, 
we shall analyze the TBA equation (\ref{tba}). 
In the second term of the right hand side of (\ref{tba}), 
we find that the factor $Y_{m}^{-1}(v)$ is not 
relevant for the calculation of the finite temperature 
correction of $T^{(1)}_{1}(v)$.  
To avoid this, we have to modify (\ref{tba}). 
After some manipulation, one can rewrite (\ref{tba}) as 
\begin{eqnarray}
\log Y_{m}(v)=
-\frac{4\pi \beta \delta_{m 1}\sinh \frac{4\pi v}{3}}
{\sqrt{3}\sinh 2\pi v}+
G*\log
\left\{\frac{(1+Y_{m+1})(1+Y_{m-1})}
{1+Y_{m}}
\right\}(v), 
\label{tba-modi}
\end{eqnarray}
where $m\in{\mathbb Z}_{\ge 1}$, $Y_{0}(v):=0$. 
We introduce the scaling function in the low temperature 
limit 
\begin{equation}
y_{m,\pm}(v)=\lim_{\beta \to \infty} Y_{m}\left(\pm
\frac{3}{2\pi}\left(v+
\log\frac{4\pi \beta}{\sqrt{3}} \right)\right).
\label{y-limit}
\end{equation}
 From the modified TBA equation (\ref{tba-modi}),
we find that $y_{m,\pm}(v)$ satisfies the 
following non-linear integral equation
\begin{eqnarray}
\log y_{m,\pm}(v)=
-{\rm e}^{-v}\delta_{m 1}+
\widetilde{G}*\log
\left\{\frac{(1+y_{m+1,\pm})(1+y_{m-1,\pm})}
{1+y_{m,\pm}}
\right\}(v), 
\label{tba-limit}
\end{eqnarray}
where $m\in{\mathbb Z}_{\ge 1}$, $y_{0,\pm}(v):=0$ and 
\begin{equation}
\widetilde{G}(v)=
\frac{3}{2\pi}G\left(\frac{3}{2\pi}v\right).
\label{kernel2}
\end{equation}
We shall divide $T^{(1)}_{1}(v)$ into 
the ground state and the finite temperature correction parts,
$T_1^{\rm gs}(v)$ and $T_1^{\rm fn}(v)$, respectively, 
\begin{eqnarray}
T^{(1)}_{1}(v)=T_{1}^{\rm{gs}}(v)T_{1}^{\rm{fn}}(v). 
\end{eqnarray}
By using the functions (\ref{y-limit}), 
one can derive the asymptotic behavior of 
the finite temperature correction part 
for large $\beta$: 
\begin{eqnarray}
\log T_{1}^{\rm fn}(v)&=&G*\log(1+Y_{1})(v) \nonumber \\ 
&=& \frac{\sqrt{3}}{\pi}
\int_{-\log\frac{4\pi \beta}{\sqrt{3}}}^{\infty}
{\rm d}w 
\frac{\sinh \frac{4\pi}{3}
\left(v-
\frac{3}{2\pi}
\left(w+\log\frac{4\pi \beta}{\sqrt{3}}\right)
\right)}
{\sinh 2\pi 
\left(v-
\frac{3}{2\pi}
\left(w+\log\frac{4\pi \beta}{\sqrt{3}}\right)
\right)} 
\nonumber \\ 
&& \hspace{60pt} \times \log \left(1+
Y_{1}\left(\frac{3}{2\pi}
\left(w+\log\frac{4\pi \beta}{\sqrt{3}}\right)\right)
\right)
\nonumber \\ 
&&+\frac{\sqrt{3}}{\pi}
\int_{-\log\frac{4\pi \beta}{\sqrt{3}}}^{\infty}
{\rm d}w 
\frac{\sinh \frac{4\pi}{3}
\left(v+
\frac{3}{2\pi}
\left(w+\log\frac{4\pi \beta}{\sqrt{3}}\right)
\right)}
{\sinh 2\pi 
\left(v+
\frac{3}{2\pi}
\left(w+\log\frac{4\pi \beta}{\sqrt{3}}\right)
\right)} 
\nonumber \\ 
&& \hspace{60pt} \times \log \left(1+
Y_{1}\left(-\frac{3}{2\pi}
\left(w+\log\frac{4\pi \beta}{\sqrt{3}}\right)\right)
\right)
\nonumber \\ 
&=& \frac{3}{4\pi^{2} \beta}
\left\{{\rm e}^{\frac{2\pi}{3}v}
\int_{-\infty}^{\infty}
{\rm d}w {\rm e}^{-w}\log(1+y_{1,+}(w)) \right.
\nonumber 
\\&& \left.+
{\rm e}^{-\frac{2\pi}{3}v}
\int_{-\infty}^{\infty}
{\rm d}w {\rm e}^{-w}\log(1+y_{1,-}(w))
\right\}
+o(\frac{1}{\beta}) .
\label{kitarou}
\end{eqnarray}
Taking account of the relation $\widetilde{G}(-v)=\widetilde{G}(v)$, 
we can derive the following relation from (\ref{tba-limit}): 
\begin{eqnarray}
&& \hspace{-20pt}
\int_{-\infty}^{\infty} {\rm d}v {\rm e}^{-v}\log(1+y_{1,\pm}(v))
=\frac{1}{2} \sum_{m=1}^{\infty}
\int_{-\infty}^{\infty} {\rm d} v
\left\{
\log(1+y_{m,\pm}(v))\frac{{\rm d}}{{\rm d}v}
\log y_{m,\pm}(v) \right.
\nonumber \\ 
&& \hspace{160pt}
\left.
-\log y_{m,\pm}(v)\frac{{\rm d}}{{\rm d}v}\log(1+y_{m,\pm}(v))
\right\}
\nonumber \\ 
&& \hspace{60pt}
=\frac{1}{2} \sum_{m=1}^{\infty}
\int_{y_{m,\pm}(-\infty)}^{y_{m,\pm}(\infty)} {\rm d}y
\left\{
\frac{\log(1+y)}{y}-\frac{\log y}{1+y}
\right\} 
\nonumber \\ 
&& \hspace{60pt}=
\sum_{m=1}^{\infty}
\left\{L\left(\frac{y_{m,\pm}(\infty)}{1+y_{m,\pm}(\infty)}\right)
-L\left(\frac{y_{m,\pm}(-\infty)}{1+y_{m,\pm}(-\infty)}\right) \right\},
\label{tba-sum}
\end{eqnarray}
where we assume that $y_{m,\pm}(v)$ are non-decreasing functions
 on $v \in {\mathbb R}$ and
$L(v)$ is  the Rogers dilogarithm function  
\begin{eqnarray}
L(x)=-\frac{1}{2}\int_{0}^{x} {\rm d}y
\left\{
\frac{\log(1-y)}{y}+\frac{\log y}{1-y}
\right\}
\quad \mbox{for} \quad 0 \le x \le 1.
\label{Roger}
\end{eqnarray}
Substituting (\ref{tba-sum}) into (\ref{kitarou}), we obtain
\begin{eqnarray}
\log T_{1}^{\rm{fn}}(v)
&=& \frac{3}{4\pi^{2} \beta}
\left[{\rm e}^{\frac{2\pi}{3}v}
\sum_{m=1}^{\infty}
\left\{L\left(\frac{y_{m,+}(\infty)}{1+y_{m,+}(\infty)}\right)
-L\left(\frac{y_{m,+}(-\infty)}{1+y_{m,+}(-\infty)}\right) \right\}
 \right.
\nonumber 
\\ && \hspace{-60pt}
\left.+
{\rm e}^{-\frac{2\pi}{3}v}
\sum_{m=1}^{\infty}
\left\{L\left(\frac{y_{m,-}(\infty)}{1+y_{m,-}(\infty)}\right)
-L\left(\frac{y_{m,-}(-\infty)}{1+y_{m,-}(-\infty)}\right) \right\}
\right]
+o(\frac{1}{\beta}).
\end{eqnarray}
In our case, both $y_{m,+}(v)$ and $y_{m,-}(v)$ behave in 
the same manner, 
thus we set $y_{m}(v):=y_{m,\pm}(v)$. 
For small $T$, the leading term of the specific heat $C$
\begin{eqnarray}
C=
-\frac{\partial}{\partial T}
\left(T^{2}\frac{\partial}{\partial T} 
\left(\frac{f}{T}\right) \right)
=
\frac{\partial}{\partial T}
\left(T^{2}\frac{\partial}{\partial T} \log T_{1}^{\rm {fn}}(0) \right)
\label{specific}
\end{eqnarray}
is proportional \cite{BCN86,A86} to 
the central charge c: 
\begin{eqnarray}
C=\frac{\pi c T}{3v_{\rm F}} + o(T), \label{katochan}
\end{eqnarray}
where the Fermi velocity is $v_{\rm F}=2\pi/3$ \cite{Mar95}. 
 Thus, we can express the central charge  as 
\begin{eqnarray}
c&=& 
\frac{6}{\pi^{2}}
 \sum_{m=1}^{\infty}
\left\{L\left(\frac{y_{m}(\infty)}{1+y_{m}(\infty)}\right)
-L\left(\frac{y_{m}(-\infty)}{1+y_{m}(-\infty)}\right) \right\}.
\label{central}
\end{eqnarray}
Above expression is widely seen in the model related
 to rank one algebras 
(see for example, refs. \cite{BR89,KP92,K93,KNS94-2}). 

Now we shall evaluate the limit $y_{m}(\pm \infty)$. 
For $v \to \infty$, (\ref{tba-limit})
 reduces to the constant $Y$-system\cite{ST99}
\begin{eqnarray}
y_{m}=
\frac{(1+y_{m+1})(1+y_{m-1})}{1+y_{m}} 
\quad \mbox{for} \quad m \in {\mathbb Z}_{\ge 1}, 
\label{Y-sys}
\end{eqnarray}
where $y_{0}:=0$. 
Thus, we expect 
(see, eq. (4.1) in ref. \cite{ST00}) 
\begin{eqnarray}
\lim_{v \to \infty} y_{m}(v)=\frac{m(m+3)}{2}.
\label{limit+}
\end{eqnarray}
The divergence from the 
first term in rhs of eq. (\ref{tba-limit}) 
in the limit $v \to -\infty$ 
is expected to be 
canceled by lhs if 
\begin{eqnarray}
  y_{1}(v) \longrightarrow +0 \qquad \mbox{for} \quad v \to -\infty .
\end{eqnarray}
Then remaining $y_{m}(v)$ for $m \in {\mathbb Z}_{\ge 2}$ 
obey the constant $Y$-system (\ref{Y-sys}). 
Thus we expect $y_{m}(-\infty)$ is given as the 
solution of (\ref{Y-sys}) with $m \to m-1$:
\begin{eqnarray}
\lim_{v \to -\infty} y_{m}(v)=\frac{(m-1)(m+2)}{2} 
\quad {\rm for} \quad m \in {\mathbb Z}_{\ge 2}.
\label{limit-}
\end{eqnarray}
Using these relations (\ref{central}), (\ref{limit+}), 
(\ref{limit-}) and 
 the fact $L(1)=\pi^2/6$, we finally arrive at $c=1$. 
This agrees with the conjecture \cite {Mar95} 
by the root density method. 

%
Now we shall solve the TBA equation (\ref{tba-modi}) 
numerically to analyze the thermodynamic quantities
 at finite temperatures. 
 To analyze this equation (\ref{tba-modi}) numerically, 
 one needs to truncate it as a finite number 
 (we call this number $m_t$) of nonlinear integral equations 
since it is composed of an infinite number of 
nonlinear integral equations.
The validity of this approximation can be verified 
from the fact that the numerical results are almost independent 
of $m_t$ when $m_t$ is more than a certain large 
number.

\begin{figure}
\begin{center}
\includegraphics[width=0.49\textwidth]{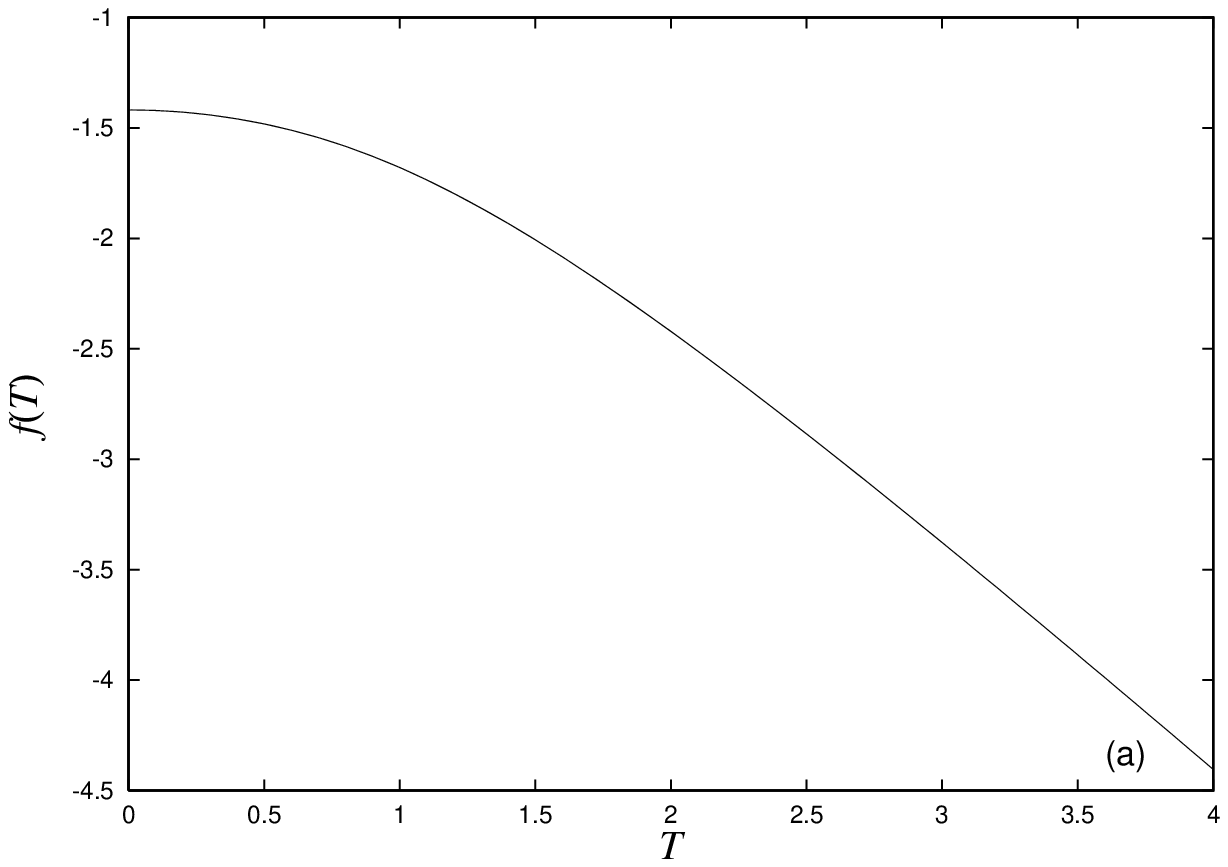}
\includegraphics[width=0.49\textwidth]{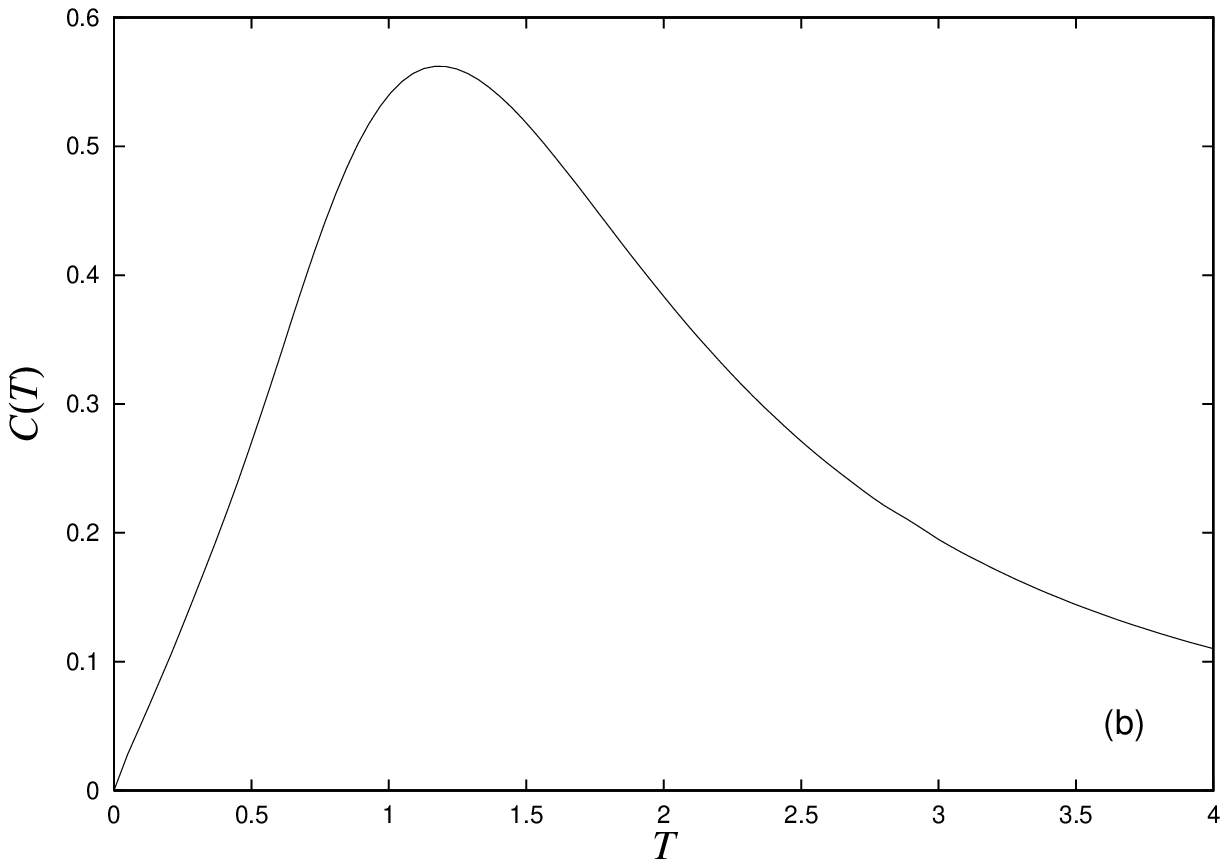}
\includegraphics[width=0.49\textwidth]{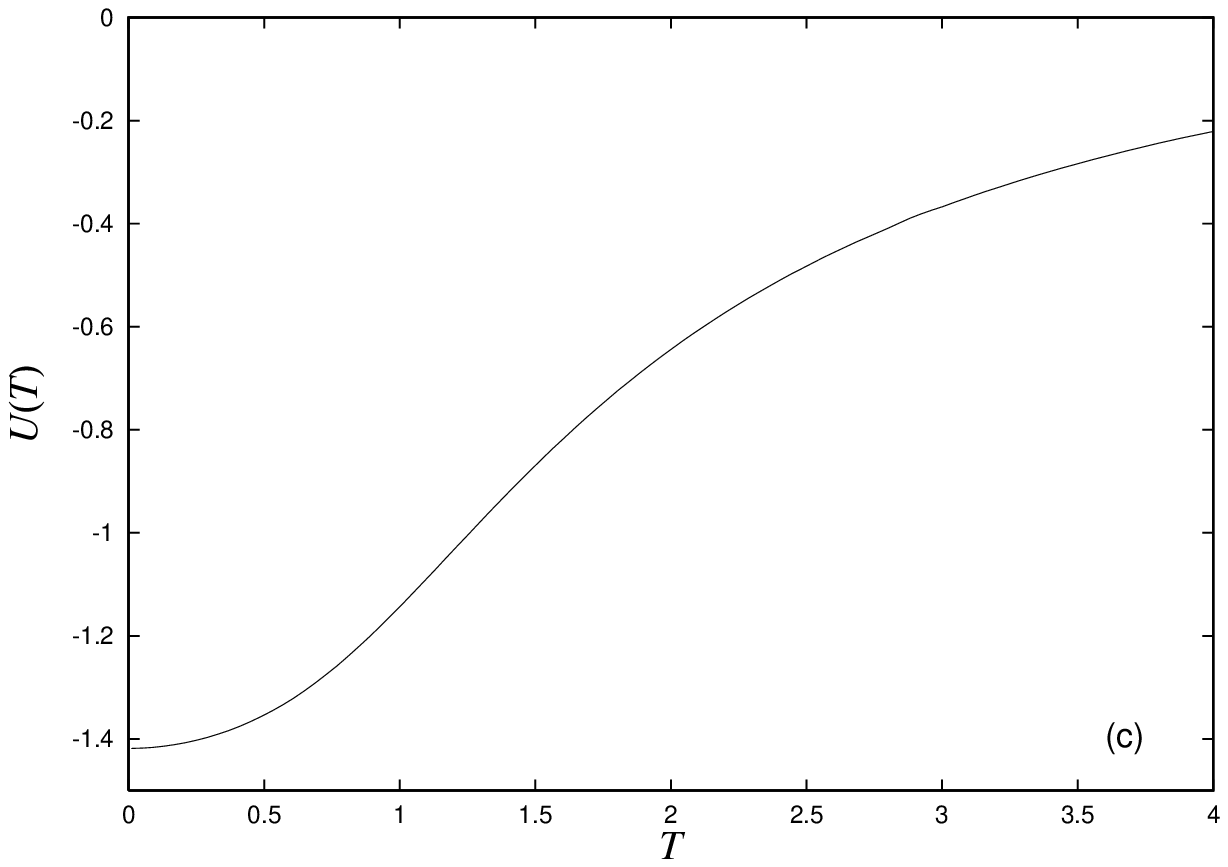}
\includegraphics[width=0.49\textwidth]{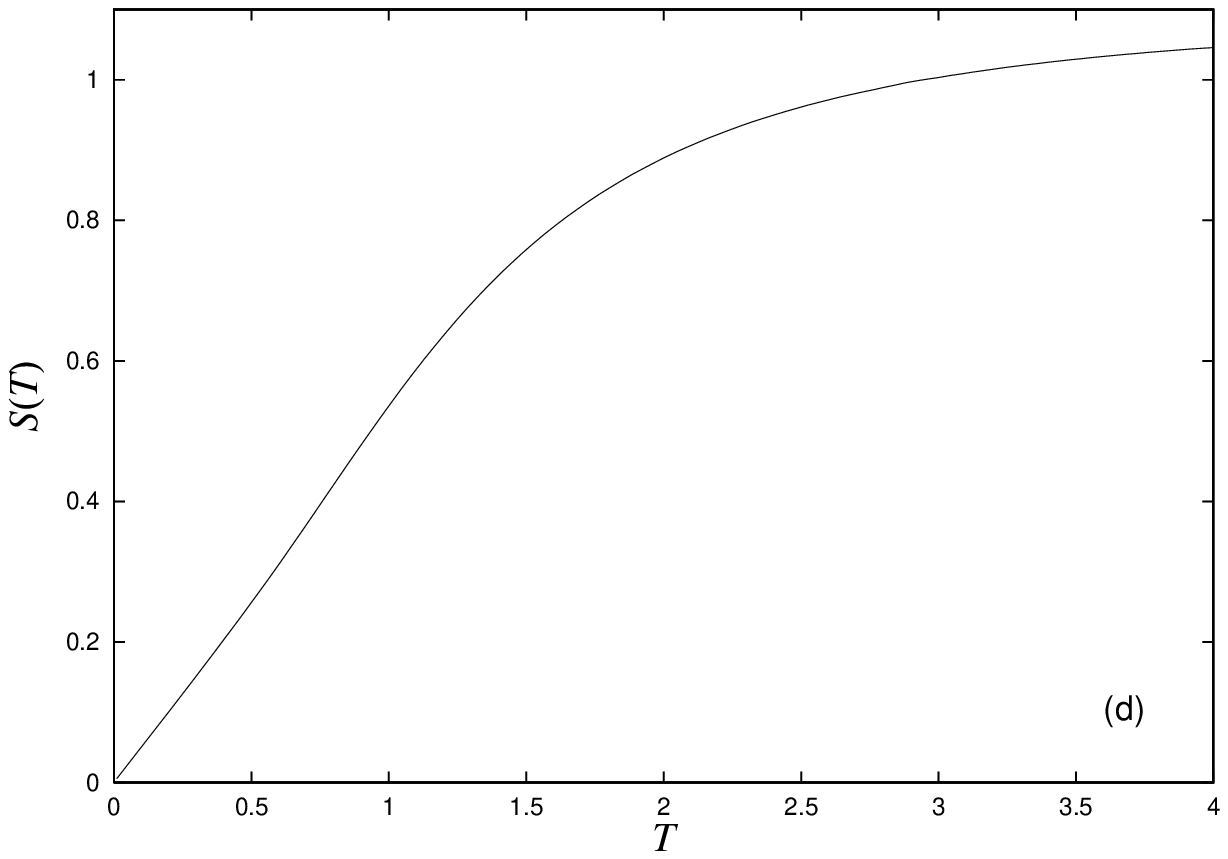}
\end{center}
\caption{Temperature dependence of fundamental 
thermodynamic quantities: (a) Free energy $f(T)$, (b)
Specific heat $C(T)$, (c) Internal energy $U(T)$ and
(d) Entropy $S(T)$. }
\label{fig}
\end{figure}

Figure \ref{fig} is numerical results of temperature dependence
of some important thermodynamic quantities such as
the free energy  (\ref{free2}),  the specific heat (\ref{specific}),
the entropy $S=\frac{\partial}{\partial T} (T\log T_1^{(1)}(0))$ and
the internal energy $U=T^2 \frac{\partial}{\partial T}(\log T_1^{(1)}(0))$.
These results agree with 
well-known values at the both special limits: 
(i) at the low temperature limit, 
the free energy and the internal energy
agree with the ground state energy \cite{Mar95} 
$E_{\rm gs}\approx -1.4184$ 
and the specific heat shows proportionate increase with 
respect to the temperature,  which leads the central 
charge $c=1$, (ii) at the high temperature limits, the
entropy agrees with the value $\log 3 \approx 1.09861$ derived 
from the fact that the present model has three states.
\section{Comment on the Excited State}
The excited state TBA equations 
 are characterized by 
 zeros of DVF $T^{(1)}_{m}(v)$ of the fused QTM 
 and a phase factor. 
 They have the following form (cf. ref. \cite{ST00}). 
\begin{eqnarray}
&&
\hspace{-20pt} \log Y_{m}(v)=
-\frac{\pi \beta \delta_{m 1}}{\cosh\pi v}+
K*\log
\left\{\frac{(1+Y_{m+1})(1+Y_{m-1})}
{1+Y_{m}^{-1}}
\right\}(v) \nonumber \\ 
&&  \hspace{50pt} +\log \left\{ 
\frac{\prod_{\{z_{m+1}\}} \tanh \frac{\pi}{2}(v-z_{m+1})
\prod_{\{z_{m-1}\}} \tanh \frac{\pi}{2}(v-z_{m-1})}
{\prod_{\{z_{m}\}} \tanh \frac{\pi}{2}(v-z_{m})}
\right\}
\nonumber \\ 
&& \hspace{50pt} +
\{1-(-1)^{m}\}\{1-(-1)^{\zeta_{\infty}}\}\frac{\pi i}{4}, 
\label{tba-ex}
\end{eqnarray}
where $m\in{\mathbb Z}_{\ge 1}$; $Y_{0}(v):=0$; 
$\zeta_{\infty}=\lim_{N \to \infty}(N-n)$; 
$N$: the Trotter number; $n$: a quantum number in the 
Bethe ansatz equation (BAE); 
$\{z_{m}\}$ are zeros of $T^{(1)}_{m}(v)$ in the physical strip 
${\rm Im} v \in [-1/2,1/2]$, which can also be characterized as
\begin{equation}  
Y_{m}(z_{m}\pm \frac{i}{2})=-1.
\label{sub} 
\end{equation}
For the second largest eigenvalue case 
($n=N-1$, $\{z_{m}\}=\{\pm x_{m}|m \in {\mathbb Z}_{\ge 1} \}$: 
there is a misprint
\footnote{This misprint was corrected in 
math-ph/9912014 {\bf v3}.} in the phase factor 
of eq.(5.3) in ref. \cite{ST00}.), 
we have to consider zeros of $T^{(1)}_{m}(v)$, 
which enter into the physical strip 
for all $m\in {\mathbb Z}_{\ge 1}$.  
In addition, the second largest eigenvalue 
is conjectured to be described by the same distribution pattern 
(see, Fig.3 and Fig.5 in ref. \cite{ST00})
of the root of the BAE at any finite temperatures. 
On the other hand, for the third largest eigenvalue case, 
 our numerical study for finite $N$ indicates that the  
distribution pattern of the roots of the BAE 
 may vary with temperature. 
This suggests a possibility of the occurrence of the level crossing. 
We have also observed numerically that 
only the zeros of $T^{(1)}_{1}(v)$  
appear in the physical strip for one of the distribution 
 patterns for the third largest eigenvalue. 

Numerical studies about such excited states are also
a crucial problem. 
However to solve the excited state TBA equation, 
we have to analyze not only infinite number of unknown 
functions but also their zeros which are determined by 
subsidiary condition (\ref{sub}).  
To make numerical results converge, one needs
to determine all the zeros in the physical strip accurately. 
For this reason the truncation approach available 
for solving the TBA equation have no use for 
the excited  state case.
%
\section*{Acknowledgments}
\noindent
The authors would like to thank Professor J. Suzuki
 for useful suggestions. 
KS thanks Professor A. Kl\"umper for comments. 
KS acknowledges financial support by the {\it 
Deutsche Forschungsgemeinschaft} under grant 
No. Kl645/3-3.  
ZT is supported by JSPS Research Fellowships for Young Scientists. 
\eqreset
\renewcommand{\theequation}{A.1.\arabic{equation}}
\section*{Appendix:
 Finite Temperature Correction from the String Hypothesis}
We can also confirm the central charge $c=1$ 
  based on the string hypothesis 
  (see, refs. \cite{B83,TW83,BR89,K93}). 
In this case, the $Y$-function in TBA (\ref{tba-modi}) 
corresponds to the ratio of the
particle density $\rho_{m}^{\rm p}(v)$ and the 
hole density $\rho_{m}^{\rm h}(v)$  
: $Y_{m}(v)=\rho_{m}^{\rm h}(v)/\rho_{m}^{\rm p}(v)$.
{}From our previous results (eq. (15) in ref. \cite{ST99}),
the particle and hole densities satisfy the  following 
relation. 
\begin{equation}
\frac{2\delta_{m1} \sinh\frac{4\pi v}{3}}{\sqrt{3}\sinh 2\pi v}=
\rho_m^{\rm p}(v)+\rho_m^{\rm h}(v)+G*\rho_{m}^{\rm h}(v)-
G*\rho_{m-1}^{\rm h}(v)-G*\rho_{m+1}^{\rm h}(v),
\label{density}
\end{equation} 
where we define $\rho_0^{\rm h}=0$. 
To consider the low temperature behavior of 
(\ref{tba-modi}), we set $Y_{m}(v)=\exp(\beta \epsilon_m(v))$ 
 and shift the variable $v \to \frac{3}{2\pi}(v+
\log\frac{4\pi \beta}{\sqrt{3}})$.
For $T\ll 1$, 
(\ref{tba-modi}) can be expressed in terms of the function 
$\varphi_{m}(v):=
\beta\epsilon_m(\frac{3}{2\pi}(v+\log\frac{4\pi \beta}{\sqrt{3}}))$,
\begin{eqnarray}
e^{-v}\delta_{m1}&=&
\log(1+\exp(-\varphi_m(v)))-\log(1+\exp(\varphi_m(v))) \nonumber \\
&&-\widetilde{G}*\log(1+\exp(\varphi_m))(v)+
\widetilde{G}*\log(1+\exp(\varphi_{m-1}))(v) 
\nonumber \\
&&+\widetilde{G}*\log(1+\exp(\varphi_{m+1}))(v).
\label {lowmtba}
\end{eqnarray}
This is effectively the same equation as (\ref{tba-limit}). 
Requiring the same shift for (\ref{density}) and
comparing the result with the $v$-derivative of (\ref{lowmtba}),
we find the asymptotics  of the particle
and hole densities in the limit $v\to\infty$: 
\begin{eqnarray}
&&\rho^{\rm p}_{m}(v) \simeq \frac{3}{4\pi^2}f(\beta \epsilon_m(v))
\frac{{\rm d}}{{\rm d}v}\epsilon_m(v) \nonumber \\
&&\rho^{\rm h}_{m}(v)\simeq \frac{3}{4\pi^2}(1-f(\beta \epsilon_m(v)))
\frac{{\rm d}}{{\rm d}v}\epsilon_m(v),
\label{densityasym}
\end{eqnarray}
where $f(\varphi)=(1+e^{\varphi})^{-1}$ is the Fermi
distribution function.
Using  (\ref{densityasym}) and taking into account the
contribution from the negative large $v$ 
derived by the shift 
 $v\rightarrow -\frac{3}{2\pi}(v+\log\frac{4\pi \beta}{\sqrt{3}})$,
we find the entropy per site (eq. (21) in ref. \cite{ST99}) 
have the following asymptotics 
\begin{eqnarray}
S&=&-\frac{3}{2\pi^2 \beta}\sum_{m=1}^{\infty}
\int_{\varphi_m(-\infty)}^{\varphi_m(\infty)} {\rm d}\varphi 
\left\{
f(\varphi)\log f(\varphi)+(1-f(\varphi))\log(1-f(\varphi))
\right\} \nonumber \\ 
&=& \frac{3}{\pi^2 \beta}\sum_{m=1}^{\infty}
\left\{
L(1-f(\varphi_{m}(\infty)))-
L(1-f(\varphi_{m}(-\infty)))
\right\}.
\end{eqnarray}
{}Using the fact 
$C=T(\partial S/\partial T)$, eq. (\ref{katochan}) and the identification 
$y_{m}(v)=\lim_{T \to 0}\exp(\varphi_{m}(v))$,
we can reconfirm the central charge $c=1$.
  
\end{document}